\documentstyle[12pt,epsfig]{article}
\def\bge{\begin{equation}}
\def\ene{\end{equation}}
\def\bg{\begin{eqnarray}}
\def\en{\end{eqnarray}}

\begin{document}
%%%%%%%%%%%%%%%%%%%%%%%%%%%%%%%%%%%%%%%%%%%%%%%%%%%%%%%%%%%%%%%%%%%%%%%%
\renewcommand{\thefootnote}{\fnsymbol{footnote}}
\begin{flushright}
ADP-99-32/T369 \\
IU/NTC 99-04
\end{flushright}
\begin{center}
{\LARGE A New Analysis of Charge Symmetry Violation \\ 
in Parton Distributions}
\end{center}
\vspace{0.5cm}
\begin{center}
\begin{large}
C.~Boros$^1$\footnote{cboros@physics.adelaide.edu.au}, 
F.M.~Steffens$^2$\footnote{fsteffen@if.usp.br},
J.T.~Londergan$^3$\footnote{tlonderg@iucf.indiana.edu} and 
A.W.~Thomas$^1$\footnote{athomas@physics.adelaide.edu.au}
\\ 
\end{large} 
$^1$Special Research Center for the Subatomic Structure of Matter \\
and Department of Physics and Mathematical Physics \\
The University of Adelaide, SA 5005, Australia 
\\
$^2$Instituto de Fisica, USP, C. P. 66 318, 05315-970,SP, Brasil 
\\
$^3$Department of Physics and Nuclear Theory Center \\
Indiana University, Bloomington, IN 47408, USA
\end{center}
%
%\vspace{0.5cm}
%\newpage
%
\begin{abstract}
To date, the strongest indication of charge symmetry violation 
in parton distributions has been obtained by comparing 
the $F_2$ structure functions from 
CCFR neutrino data and NMC muon data. We show that in order to 
make precise tests of charge symmetry with the neutrino data, two 
conditions must be satisfied. First, the nuclear shadowing 
calculations must be made explicitly for
neutrinos, not simply taken from muon data on nuclei. Second, the
contribution of strange and charm quarks should be calculated explicitly
using next-to-leading order [NLO] QCD, and the ``slow rescaling''
charm threshold correction should
not be applied to the neutrino data. When these criteria are satisfied,
the comparison is consistent with charge symmetry within the
experimental errors and the present uncertainty in the strange quark
distribution of the nucleon.
\end{abstract}
%\vspace{0.5cm}
%PACS numbers: \\
%Keywords:  
%
%%%%%%%%%%%%%%%%%%%%%%%%%%%%%%%%%%%%%%%%%%%%%%%%%%
\newpage

In recent years there have been a number of surprising discoveries
concerning the parton distributions of the nucleon. In the valence
distributions it now seems likely that the $d/u$ ratio at large-$x$ is
consistent with the expectations of perturbative QCD \cite{doveru}, 
when for many
years it seemed that the ratio might vanish in that region. However, it
is in the sea that one has found the biggest surprises. Following the
discovery of a violation of the Gottfried sum-rule by the New
Muon Collaboration [NMC] \cite{NMC}, the E866 experiment  
at Fermilab has mapped out a clear violation of the perturbative QCD
expectation of equal numbers of $\bar{d} d$ and $\bar{u} u$ pairs in the 
sea of the proton \cite{SU2F}. Most recently, a careful comparison of
the $F_2$ structure functions measured in $\nu$ reactions by 
the CCFR Collaboration \cite{CCFR} 
and $\mu$ deep inelastic scattering by the NMC group\cite{NMC_D}, 
has revealed \cite{Bor98} 
a discrepancy that suggests a violation of
charge symmetry in the parton distributions of the nucleon, at a level
significantly larger than the expectations of either theory 
\cite{CSV_papers} or other experiments \cite{W_asym}.

Charge symmetry, which is related to the invariance of the strong
Hamiltonian under rotations of 180 degrees about the 2-axis in 
isospace \cite{CS_reviews}, implies the equality of the $d$-distribution
in the proton, $d^p$, and the $u$-distribution in the neutron, $u^n$,
etc. Charge symmetry implies analogous relations for the antiquark
distributions. It is implicit in the standard notation for 
parton distributions,
with $d \equiv d^p = u^n, u \equiv u^p = d^n$, and so on.
In view of the importance of any significant violation of 
charge symmetry it is vital to reexamine every aspect of the analysis of
the CCFR and NMC data and, in particular, the 
 difference 
 \begin{equation}
 \frac{5}{6} F_2^{\nu N}(x,Q^2) - 3 F_2^{\mu N}(x,Q^2) ~~, 
\label{eqn1}
\end{equation}
on an isoscalar target -- after corrections for $s$ and
$c$ quarks, this difference should be strictly zero for all 
$x$ if charge symmetry is exact.

The comparison between the $\nu$ and $\mu$ data is complicated by the
fact that the $\nu$ data are taken on an Fe target. The heavy 
target is necessary because of the low
event rates for neutrino experiments. Before these data can be compared
with the muon data taken on deuterium, a number of corrections must be
made. In what follows we re-examine all of these corrections. We 
can now make rather accurate estimates of these corrections 
as a result of the rapid
experimental \cite{heavy_expt} and theoretical developments
\cite{gkr,acot,buza98,thorne98,heavy_theory}  
in our understanding of charm structure functions. In particular, 
we have applied a next-to-leading order [NLO] calculation of the 
charm quark contribution \cite{heavy_theory}
to both the $\nu$ and $\mu$ experiments, rather
than applying a ``slow rescaling'' correction to the data, as has been
the custom \cite{CCFR,Seligman}.  In this way we can compare 
our theoretical NLO calculation directly
with the data.

\begin{figure}[htb]
\begin{center}
\epsfig{file=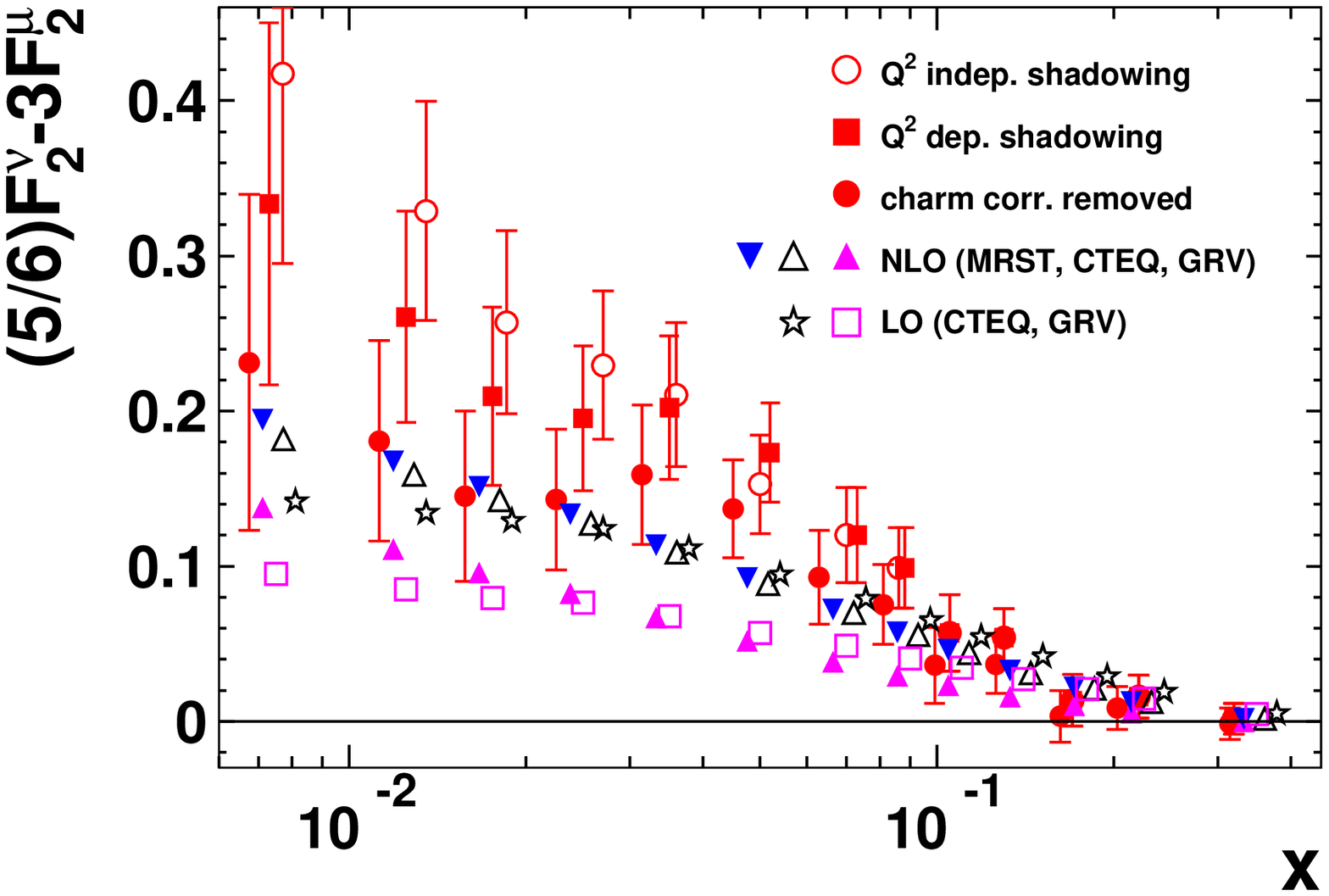,height=10cm}
\caption{Comparison between theory and experiment for the difference
$\frac{5}{6} F_2^{\nu} - 3 F_2^{\mu}$, which is sensitive to deviations
from charge symmetry in the parton distributions. The open circles use 
the original CCFR data, where the nuclear corrections to the 
$\nu$ data are taken from
muon measurements \protect\cite{CCFR}. The solid squares involve the
same data, but the shadowing corrections have been made explicitly for
neutrinos \protect\cite{Bor98,Shadow}. The solid circles are the same as
the solid squares except that the ``slow rescaling'' correction has been
removed. The open squares, stars and the  triangles are respectively 
LO (massless)  and NLO calculations,   
including charm mass effects \protect\cite{heavy_theory} 
and using different parametrizations for the parton distributions.  
(Note that the theoretical calculations are all made at the {\em same}
$x$ and $Q^2$ as the data, but displaced slightly for clarity.)
}
\label{fig1}
\end{center}
\end{figure}

We shall see that, when {\it both} the $\nu$ shadowing corrections
\cite{Shadow} and the explicit charm production calculations are made
with the best available theory, the residual discrepancy between 
the $\nu$ and $\mu$ data is much smaller than was suggested 
by earlier analysis. Our key results are summarized in Fig. 
\ref{fig1}, where the NLO results (triangles) should be 
compared with the data (filled circles).
The difference $(5/6)F_2^{\nu N} - 3 F_2^{\mu N} \simeq
(s + \overline s)/2$ is quite sensitive to the strange distribution
in the proton.  Consequently, the agreement between experiment 
and theory will depend upon the particular parton distribution 
which is used.  The data is still systematically above theoretical 
expectations based on the GRV98 \cite{grv98} parametrization and on 
charge symmetry in the region $0.008 \leq x \leq 0.08$ (these 
are the solid gray triangles in Fig.\ \ref{fig1}).  
However, the discrepancy is no longer outrageously large, as was 
suggested by earlier analysis.  If one uses the MRST99 
\cite{mrst} or CTEQ5 \cite{cteq} parton distributions for the 
NLO calculation of the structure functions (these are 
respectively the solid inverted triangles and the open triangles
in Fig.\ \ref{fig1}), then the theory and experiment are in very 
good agreement.  Both the MRST and CTEQ theoretical predictions are 
within one standard deviation of the experimental points 
for all $x$.

Our purpose here is to explain the meaning of the various theoretical
and experimental points in Fig. \ref{fig1}, leading to our final
conclusion concerning the level of charge symmetry violation consistent
with the data. We begin with the corrections which are well under
control. 

Since Fe is not an isoscalar target, the structure function 
extracted from $\nu$-Fe scattering  
has to be converted to that of an isoscalar target. This is done 
by correcting for the excess neutrons. This correction is 
relatively small and straightforward. The quantity  
$5/6F_2^{\nu N} - 3 F_2^{\mu N}$ contains contributions from both strange 
quarks and charge symmetry violation [CSV].  CSV tests which 
involve comparisons of $F_2$ structure functions from muon and 
neutrino scattering are quite sensitive to the constraints 
on the strange quark distributions. The strange quark distribution 
has been measured independently 
in dimuon production  in neutrino DIS by the CCFR 
Collaboration \cite{CCFRstrange}. It is about a factor of two smaller 
than what is needed to resolve the original CCFR-NMC discrepancy. 
The difference, $5/6F_2^{\nu N} - 3 F_2^{\mu N}$, is also sensitive  
to the difference between the strange and anti-strange 
quark distributions, $xs(x)-x\bar s(x)$. However, our previous analysis 
\cite{Bor98} has shown that a {\it negative} anti-strange 
quark distribution 
would be necessary to account for the whole  
NMC-CCFR discrepancy. We will come back to the uncertainties 
related to the strange quark distribution after 
having discussed the nuclear corrections. 

\noindent
{\bf Shadowing for neutrinos}

The difference between the NMC and CCFR structure functions   
is very sensitive to the nuclear corrections as can be seen 
in Fig. \ref{fig1}.  
Nuclear corrections for neutrinos (nuclear EMC effect, shadowing 
and anti-shadowing) are generally applied 
using phenomenological correction factors derived from 
charged lepton reactions. 
However, there is no reason that nuclear corrections for 
neutrinos and charged leptons should be identical.  
Since the original discrepancy between NMC and CCFR 
is significant in the small $x$ region  
where nuclear shadowing is the dominant nuclear effect, 
we re-examined shadowing corrections in neutrino DIS. 
A detailed discussion can be found in Ref.\ \cite{Shadow}. Here, we 
summarize the main results. 

We used a two phase model 
which has been successfully applied to the description of shadowing
in charged lepton DIS \cite{Badel88,Melni93}. 
In generalizing this approach to weak currents, subtle
differences between shadowing in neutrino and charged lepton DIS
arise because of the partial conservation of axial currents (PCAC)
and the coupling of the weak current to both vector and axial
vector mesons. For the axial current,
PCAC requires that shadowing
in neutrino scattering for low $Q^2$($\approx m_\pi^2$)
is determined by the absorption of pions on
the target \cite{Adler}, while at larger $Q^2$-values
axial vector mesons (e.g. the $a_1^+$ for the $W^+$ charged current) 
become important. For the
weak vector current one must include the $\rho^+$ vector meson.
Since the coupling constants are related by
$f^2_{\rho^+} =f_{a_1}^2=2f^2_{\rho^0}$, the $a_1$ 
component is suppressed because of the larger $a_1$ mass, 
and since the neutrino structure function is about a factor of four 
larger than the muon one at small $x$,  
the {\it relative} shadowing arising from VMD in neutrino DIS is 
roughly a factor of two smaller than in charged lepton DIS. 
For large $Q^2$-values,
shadowing due to Pomeron exchange (which is of leading twist)
becomes dominant, leading to identical
(relative) shadowing in neutrino and charged lepton DIS. 

In Fig. \ref{fig1} the solid squares show the difference between 
the neutrino and muon structure functions,  
corrected for shadowing using the explicit calculation for 
neutrinos. 
This is to be compared with the corresponding difference when 
a $Q^2$-independent correction obtained from charged lepton DIS 
is used (open circles). 
There are differences between the two results in the small 
$x$-region, where $Q^2$ is relatively small and  
the vector meson component plays a significant role. 
Here, the $Q^2$ dependence 
of shadowing is also important. (Remember that the 
$x$-dependence of the shadowing corrections 
in charged lepton DIS, measured in fixed target experiments,   
is strongly correlated with their $Q^2$-dependence. In order to avoid
very large shadowing corrections, we have made a cut on the
data shown in Fig. 1 to remove all experimental points with 
$Q^2 < 3$ GeV$^2$.) 

We see that shadowing corrections made 
explicitly for neutrinos 
remove part of the original discrepancy between the NMC and CCFR data. 
In a LO analysis, 
the corrected data points (solid squares) 
should be compared to the LO result shown as open boxes.  
Such a comparison rests on the assumption that 
charm mass and charm threshold corrections have been taken 
into account properly by the slow rescaling 
corrections which have been applied to the data \cite{CCFR,Seligman}.  
This was the procedure adopted in our previous analysis \cite{Bor98}.  
As we have already noted, this approximation is no longer necessary, so
in this paper, we perform a next-to-leading order analysis, 
which is summarized in the following section. 

\noindent
{\bf Structure Function Calculations}

There are two options for calculating the contribution of the
charm quark to the $F_2$ structure function for neutrino
and muon deep inelastic scattering. 
One is to use the ACOT scheme \cite{acot}, which changes
the number of active flavors of the theory by introducing a charm
distribution function. The other is to keep the number
of active flavors fixed\footnote{This scheme is generally
referred to as the Fixed Flavor Number Scheme - FFNS.}, 
in which case the charm quark
contribution to $F_2$ is calculated from the interaction between the
probe and the gluons and light quarks in the target. 
In the neutral current
case, the lowest order contribution is given by the boson - gluon
fusion, which is an ${\cal O}(\alpha_s)$ process. In the charged 
current case, the process $W^+ s \rightarrow c$, which is of 
order ${\cal O}(\alpha_s^o)$,
is the dominant contribution. 

By definition, both schemes are equivalent in the region where
$Q^2$ is not much bigger than the square of the charm quark mass, $m_c^2$. 
When $Q^2 >> m_c^2$, the ACOT scheme is superior to the FFNS because
of the large logarithms in $Q^2/m_c^2$ which appear in the partonic
boson-gluon cross section. The ACOT scheme overcomes this problem
by resumming the large logarithms in $m_c^2$ through the introduction
of a charm distribution. In practice, however, the implementation of the
ACOT scheme in $\nu$ DIS is only an additional complication. 
First, because the $Q^2$ of the CCFR data in the small $x$ 
region is not much larger 
than $m_c^2$. Second, the introduction
of a charm distribution, and of subtraction terms to the boson-gluon
cross section, will require the introduction of the $W^- c \rightarrow s$
terms. This makes the case for charm tagging in the scattering process 
difficult, as the remaining $\overline c$ anti quark (of the $\overline c c$
pair) will have to hide in the hadronic debris of the reaction. 
Hence, we will adopt the FFNS for calculating the structure 
functions. 

In the following, all the expressions for $F_2$ are written as an 
average between proton and neutron targets for the muon case,

\begin{equation}
F_2^{\mu N} (x, Q^2) = \frac{1}{2}(F_2^{\mu p}(x,Q^2) + F_2^{\mu n}(x,Q^2)),
\label{c1}
\end{equation}
and an average between proton and neutron targets and neutrino and 
anti neutrino beams in the neutrino case:
 
\begin{equation}
F_2^{\nu N} (x, Q^2) = \frac{1}{4}(F_2^{\nu p}(x,Q^2) + F_2^{\nu n}(x,Q^2)
+ F_2^{\overline \nu p}(x,Q^2) + F_2^{\overline \nu n}(x,Q^2)).
\label{c2}
\end{equation}

For 3 active flavors, and using charge symmetry at the 
quark level, we have in the FFNS:

\begin{eqnarray}
F_2^{\mu N} (x, Q^2) &=& \frac{5}{18} x [(u + \overline u 
+ d + \overline d)(x,Q^2) + \frac{2}{5}(s + \overline s) 
(x,Q^2)] \otimes C_q (x,Q^2) \nonumber \\*
&& + \frac{2}{3} g(x,Q^2)\otimes C_g (x, Q^2) + F_{2c}^{\mu N} (x,Q^2),
\label{c3}
\end{eqnarray}
with

\begin{equation}
F_{2c}^{\mu N} (x, Q^2) = e_c^2 xg(x, Q^2) \otimes H_{2g}^\mu (x, Q^2).
\label{c4}
\end{equation}
The massless quark and gluon coefficient functions, $C_q$ and $C_g$,
were calculated in \cite{bardeen,zijlstra} (in our notation, $n_f =1$
in Eq. (B.6) of \cite{zijlstra}).
The partonic cross section for the 
production of heavy quarks pairs, $H_{2g}^\mu$, was calculated 
in \cite{witten}, and can also be found in \cite{heavy_theory}. 

Similarly, in the neutrino case we have:

\begin{eqnarray}
F_2^{\nu N} (x, Q^2) &=& x \left[\frac{(1 + |V_{ud}|^2)}{2} 
(u + \overline u + d + \overline d)(x,Q^2) + |V_{us}|^2
(s + \overline s) (x,Q^2)\right] \otimes C_q (x,Q^2) \nonumber \\*
&& + 2 g(x,Q^2)\otimes C_g (x, Q^2) + F_{2c}^{\nu N} (x,Q^2),
\label{c5}
\end{eqnarray}

\begin{eqnarray}
F_{2c}^{\nu N} (x, Q^2) &=& \xi \left[\frac{|V_{cd}|^2}{2} 
(u + \overline u + d + \overline d)(\xi,\mu^2) 
+ |V_{cs}|^2 (s + \overline s) (\xi ,\mu^2)\right]\otimes 
\nonumber \\*
&& (\delta(\xi - 1) + H_{2q}^\nu (\xi,\mu^2,\lambda)) 
+ 2\xi g(\xi,\mu^2) \otimes H_{2g}^\nu (\xi,\mu^2,\lambda),
\label{c6}
\end{eqnarray}
with $|V_{ud}| = |V_{cs}| = 0.974$, $|V_{us}| = |V_{cd}| = 0.220$
and $\lambda = Q^2/(Q^2 + m_c^2)$.
The scaling variable for massive particles, in the case that 
the strange quark is treated as massless but the charm quark
as massive, is given by:

\begin{equation}
\xi = x\, \left(1 + \frac{m_c^2}{Q^2} \right).
\label{c7}
\end{equation}
The massive coefficient functions in Eq. (\ref{c6}), $H_{2q}^\nu$ and
$H_{2g}^\nu$, are the ${\cal O}(\alpha_s)$ 
corrections to the $W^+ s \rightarrow c$
process: $H_{2q}^\nu$ corresponds to the $W^+ s \rightarrow c g$ 
correction, while $H_{2g}^\nu$ corresponds to the 
$W^+ g \rightarrow c \overline s$ fusion
term. They were calculated in \cite{gottschalk}, and a factor of 
$\alpha_s(\mu^2)/2\pi$ was buried in them. Eq. (\ref{c6}) 
agrees with the corresponding expression in Ref.\cite{gkr}.

\begin{figure}[htb]
\begin{center}
\epsfig{file=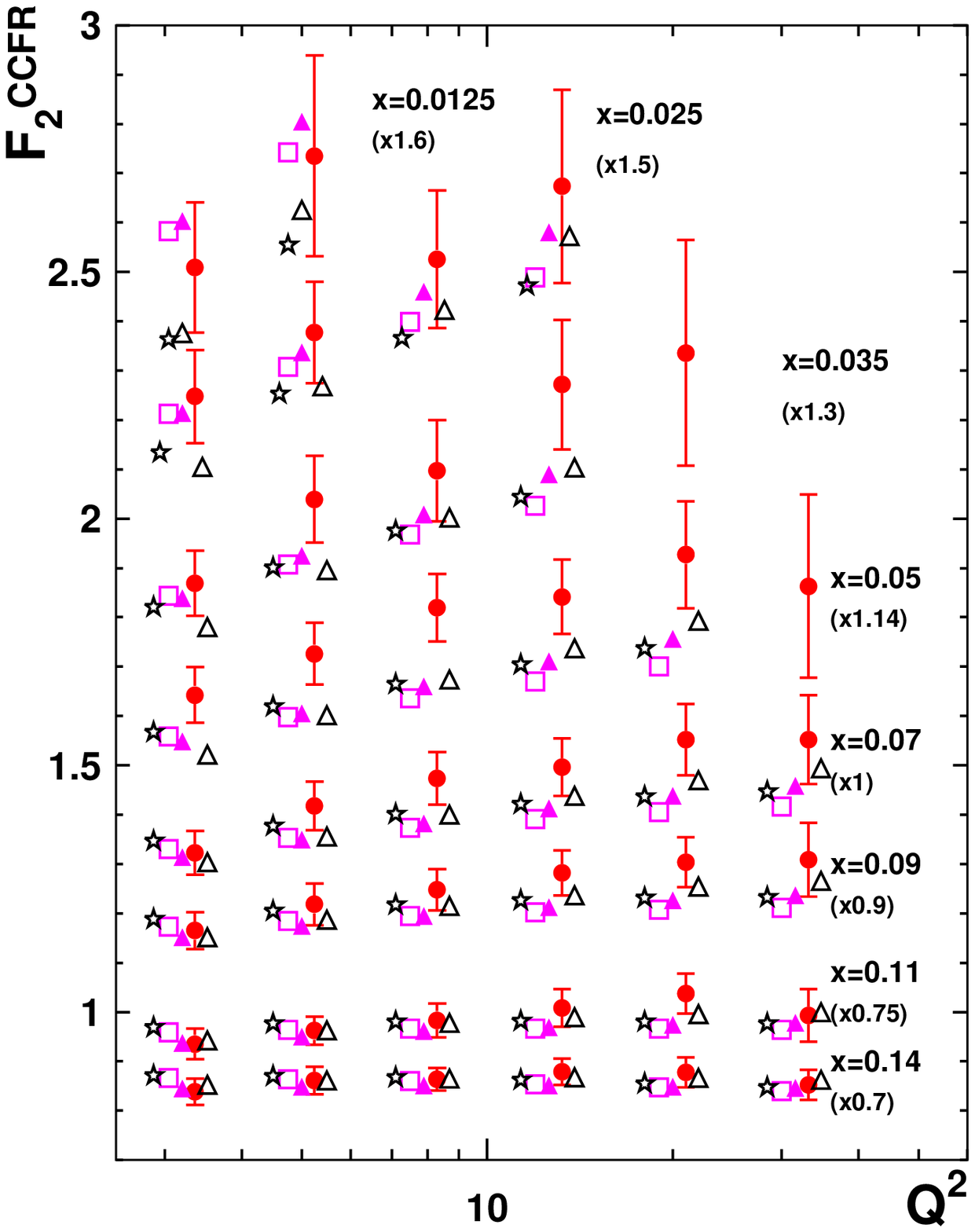,height=14cm}
\caption{The $F_2^\nu$ structure function as measured for 
neutrino charged-current DIS by the CCFR collaboration (solid 
circles), compared to LO and NLO QCD calculations.  GRV98 parton 
distributions \protect\cite{grv98}: LO = open 
squares; NLO = solid triangles.  CTEQ parton distributions 
\protect\cite{cteq}: LO = stars; NLO = open triangles.  
(Note that the theoretical calculations are all made at the {\em same}
$x$ and $Q^2$ as the data, but displaced slightly for clarity.)
The data have been corrected for shadowing, but the ``slow rescaling''
correction, present in the original CCFR data, has been removed.
}
\label{fig2}
\end{center}
\end{figure}

For the calculations, we initially used the GRV98 \cite{grv98} 
parameterization for the 3 light quarks and for the gluons, defined 
in the $\overline{MS}$ scheme. If the NMC and the CCFR data are to 
be compatible, they should be described by
the same parton distributions through Eqs. (\ref{c3})-(\ref{c6}).
Fig. \ref{fig2} shows the CCFR data points as a function of 
$Q^2$ for fixed x, against the LO (squares) and the NLO (solid 
triangles) theoretical calculations, with $\mu^2 = Q^2 + m_c^2$ for 
the factorization scale in Eq. (\ref{c6}) and $m_c = 1.4$ GeV. 
Since NLO calculations consistently
incorporate all necessary charm mass effects, 
the slow rescaling corrections have to be removed from the 
data. This is done by using the correction factors 
supplied by the CCFR Collaboration \cite{Seligman}.   
The data shown in Fig.\ \ref{fig2} have been  
corrected for nuclear effects by calculating the heavy target 
corrections specifically for neutrinos.  
The CCFR data is generally well described by both the LO and the 
NLO approaches, although the NLO calculation is superior 
for several data points.
The points at x = 0.05 and at x = 0.07, however, appear to be 
systematically below the experimental error bars.

\begin{figure}[htb]
\begin{center}
\epsfig{file=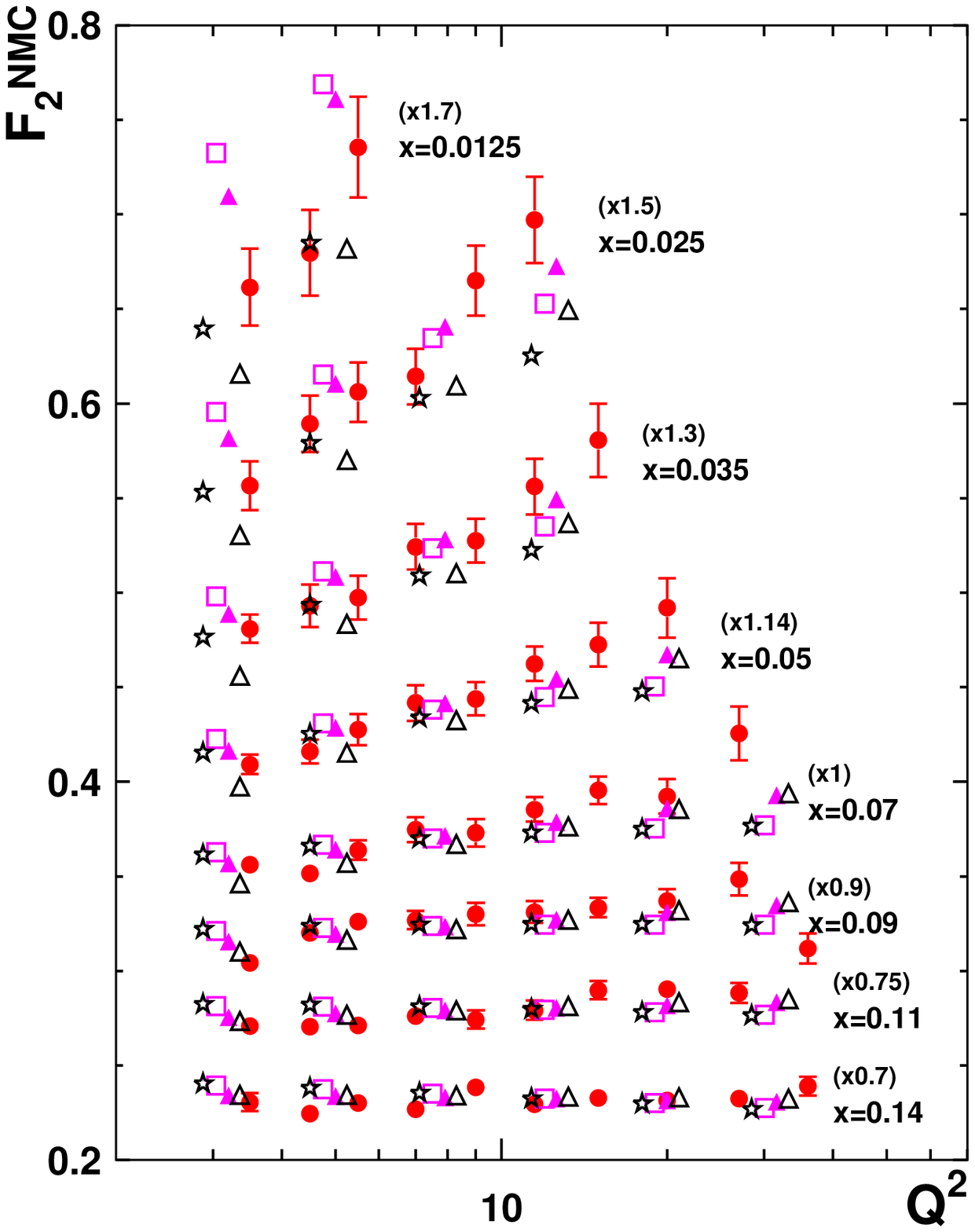,height=14cm}
\caption{The $F_2^\mu$ structure function as measured in 
muon DIS on deuterium by the NMC collaboration (solid circles), 
compared to LO and NLO QCD calculations using the GRV98 and CTEQ 
parton distributions \protect\cite{cteq,grv98}.  The notation is 
that of Fig.\ \protect\ref{fig2}.
}
\label{fig3}
\end{center}
\end{figure}

The NMC data, shown in Fig. \ref{fig3}, is also well described
by Eqs. (\ref{c3}) and (\ref{c4}), although the theory at the LO and 
at the NLO level has a slight tendency to be above the experimental data 
in the region where $x$ is small. This is an important point 
in comparing the muon and neutrino $F_2$ structure functions, 
because of the factor of 3 which
multiplies $F_2^{\mu N}$ in the difference 
$(5/6)F_2^{\nu N} - 3F_2^{\mu N}$, which is 
plotted in Fig. \ref{fig1}. 

As we mentioned previously, the results 
are dependent on the parameterization used.
To illustrate this point, we also calculated the 
structure functions $F_2^{\nu N}$ and $F_2^{\mu N}$ using the 
CTEQ \cite{cteq} (CTEQ5) parton distributions.  These are 
shown in Figs.\ \ref{fig2} and \ref{fig3}, where the LO results
are stars and the NLO results are open triangles.  The CTEQ 
parton distributions give comparable fits to the neutrino
and muon structure functions as was obtained with the GRV98 
parton distributions.  However, for the difference between the 
neutrino and muon structure 
functions given by Eq.\ \ref{eqn1}, which is plotted in Fig.\ \ref{fig1}, 
the CTEQ model gives better agreement with 
the data, as can be seen by comparing the solid circles with 
the open triangles.  The NLO calculations
with CTEQ parton distributions are essentially always within
one standard deviation of the data in Fig.\ \ref{fig1}.  We 
also carried out calculations using the MRST \cite{mrst} (MRST99) 
parton distributions \footnote{Since a LO MRST gluon distribution 
is not available,  $F_{2c}^{\mu N}$
in Eq. (\ref{c4}) has to be calculated with the NLO gluon 
distribution. However, since the difference between $F_{2c}^{\mu N}$  
calculated with the NLO MRST gluon and  
with the LO GRV98 gluon distribution is small, this approximation has 
no sizable effect on our final result.}.  The MRST fits to the 
neutrino and muon structure functions are not shown, but the 
agreement with data is comparable to that for the GRV98 and 
CTEQ models.  In Fig.\ \ref{fig1} the NLO fits using MRST are 
plotted as the inverted solid triangles.  The fit to experiment 
is very similar to that obtained with the CTEQ distributions. 

Our approach allows a fully consistent comparison  
between theory and experiment. The slow rescaling corrections   
used by the CCFR Collaboration  
are quite large, as can be seen in Fig.\ref{fig1}
(the difference between the solid circles and solid squares),  
and slow rescaling corrections play a major role in the disagreement 
between the CCFR and NMC data. 
Note that after applying the slow rescaling corrections (solid 
squares), the results should be compared to the massless  
LO calculations (open squares and stars), or to a massless NLO
calculation.  Thus the slow rescaling corrections suggest 
very large CSV contributions.  This is a completely 
different conclusion from that based on our NLO calculations, 
where the solid circles should be compared with the triangles 
in Fig.\ \ref{fig1}.    

\begin{figure}[htb]
\begin{center}
\epsfig{file=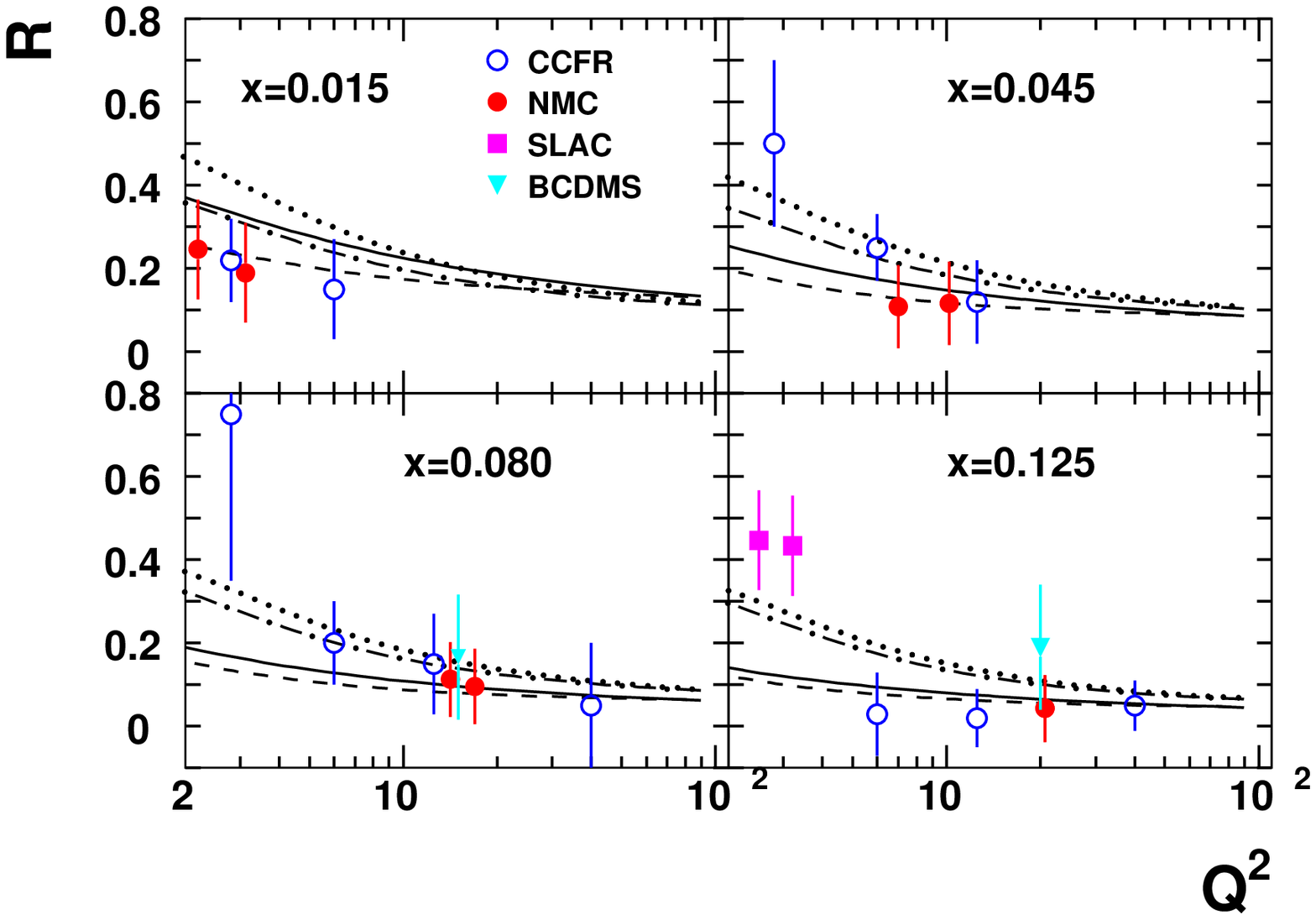,height=14cm}
\caption{The ratio of the longitudinal and transverse  
structure functions calculated in NLO for neutrino (solid line) 
and for muon (dashed line) DIS. The dash-dotted line stands 
for the parameterization of the world data on 
$R$ \protect\cite{Whitlow}. 
The dotted line is the result for $R^\nu$ using the Whitlow 
parameterization and the 
slow rescaling formalism.  
The data are from Refs. \protect\cite{Whitlow,RNMC,RBCDMS,RCCFR}. 
}
\label{fig4}
\end{center}
\end{figure}

We stress the importance of a correct treatment of the charm mass, 
along with shadowing corrections calculated specifically for 
neutrinos and not taken from muon data. These two effects allow the
experimental data to come within the range of perturbative
QCD. In the particular case of charm threshold effects, 
instead of trying
to correct the experimental data by the use of the slow rescaling
variable, we keep the data with no quark mass corrections, and
perform a QCD calculation which incorporates the effects of
a massive quark.  We also calculated the ratio of longitudinal 
and transverse structure functions both for muon ($R^\mu$) and 
neutrino DIS ($R^\nu$), using the full NLO formalism with 
all massive quark effects. The results  for $R^\nu$ and 
$R^\mu$ are shown as solid  and dashed lines, respectively,   
in Fig. \ref{fig4}. The dash-dotted line is the Whitlow 
parameterization \cite{Whitlow} of the world data 
on $R^\mu$. Both $R^\mu$ and $R^\nu$ are within the 
error bars of the available data in the kinematical region 
of our analysis ($x<0.1$ and $Q^2>3.0$ GeV$^2$) and are very close to the 
Whitlow parameterization of $R^\mu$, which was used by the 
CCFR Collaboration in the extraction of the structure functions.   
We also  calculated $R^\nu$ by assuming that $R^\nu=R^\mu$ when 
$m_c = 0$, and using the Whitlow parameterization for  
$R^\mu$. The slow rescaling variable (implemented in LO) then  
accounts for the charm mass effects, as shown in dotted line. 
We see that the differences between $R^\nu$ and $R^\mu$  
are similar in the full NLO calculation and the slow rescaling 
formalism.     
The slow rescaling correction used by the CCFR Collaboration effectively  
corrects for the  difference between $R^\nu$ and $R^\mu$ in LO.  
Removing these corrections corresponds to the structure 
functions extracted under the assumption that 
$R^\nu$ and $R^\mu$ are the same. 
Then, the difference between the experimental structure 
functions, $(5/6)F_2^\nu - 3 F_2^\mu$, will only 
involve a contribution from the difference between $R^\nu$ and $R^\mu$. 
Since the full NLO calculation correctly accounts for this 
difference, the 
calculated quantity, $(5/6)F_2^\nu - 3 F_2^\mu$, 
should be compared with the {\it uncorrected} data. 

In summary, we see that a direct comparison of the CCFR $\nu$ data and
the NMC $\mu$ data with a NLO QCD calculation leads to a much more
consistent picture, if the nuclear corrections on Fe are made specifically
for neutrinos. In order to make the comparison directly between the NLO
calculation and the data, the ``slow rescaling'' correction had to be
removed from the data. As this new analysis leads to a much less
dramatic discrepancy than earlier work, it is consistent with the
conclusion of the recent analysis of $p \bar p$ data on $W$-asymmetry
\cite{ppbar}. We observe that the data is still systematically above
the NLO calculation based on the GRV98 distributions, 
while it is in quite good agreement with calculations based on either 
the MRST or CTEQ distributions. For the present, 
the possibility of detecting any residual charge symmetry violation 
depends on resolving this uncertainty in our knowledge of the strange quark
distribution.

\vspace{1cm}
We would like to thank W. Seligman for numerous helpful communications
regarding the CCFR data. One of the authors [JTL] wishes to 
thank the Special Research Centre for the Subatomic Structure of 
Matter for its hospitality during the period when this research 
was undertaken. 
This work was supported in part by the Australian Research Council,
by FAPESP (under contracts 96/7756-6 and 98/2249-4) and by the 
National Science Foundation (research contract PHY--9722076).

\clearpage
%
%%%%%%%%%%%%% Bibliography %%%%%%%%%%%%%%%%%%%%%%%%%%%%%%%%%
%

%
\end{document}